\documentclass[aip, jmp, amsmath, amssymb, preprint]{revtex4-1}

\usepackage{tikz}
\usepackage{graphicx}
\usepackage{dcolumn}
\usepackage{bm}
\usepackage{color}
\usepackage{hyperref}
\usepackage{mathptmx}
\usepackage[title]{appendix}
\hypersetup{colorlinks   = true, urlcolor = blue, linkcolor = blue, citecolor   = blue}

\DeclareUnicodeCharacter{2009}{\,} 

\begin{document}
\title{Probing photoinduced proton coupled electron transfer process by means of two-dimensional resonant electronic-vibrational spectroscopy}
\author{Jiaji Zhang}
\email{zhang.jiaji.84e@st.kyoto-u.ac.jp}
\affiliation{Department of Chemistry, Graduate School of Science, Kyoto University, Kyoto 606-8502, Japan}

\author{Raffaele Borrelli}
\email{raffaele.borrelli@unito.it}
\affiliation{DISAFA, University of Torino,  Largo Paolo Braccini 2, I-10095 Grugliasco, Italy}

\author{Yoshitaka Tanimura}
\email{tanimura.yoshitaka.5w@kyoto-u.jp}
\affiliation{Department of Chemistry, Graduate School of Science, Kyoto University, Kyoto 606-8502, Japan}

\date{\today}

\begin{abstract}
We develop a detailed theoretical model of photo-induced proton-coupled electron transfer (PPCET) processes, which are at the basis of solar energy harvesting in biological systems and photovoltaic materials.
Our model enables to analyze the dynamics and the efficiency of a PPCET reaction under the influence of a thermal environment by disentangling the contribution of the fundamental electron transfer (ET) and proton transfer (PT) steps. In order to study quantum dynamics of the PPCET process under an interaction with non-Markovian environment we employ the hierarchical equations of motion (HEOM). 
We calculate transient absorption spectroscopy (TAS) and a newly defined two-dimensional resonant electronic-vibrational spectroscopy (2DREVS) signals in order to study the nonequilibrium reaction dynamics.
Our results show that  different transition pathways can be separated by TAS and 2DREVS.
\end{abstract}


\maketitle

\section{Introduction}

The simultaneous transfer of protons and electrons plays an important role in many natural and artificial energy conversion processes. 
A typical example is the oxygen evolving complex (OEC) of natural photosynthetic system, where the oxygen generation consists of four stepwise proton-coupled electron transfer (PCET) catalyzed reactions. 
{\cite{A.Migliore.CR.2014, J.D.Megiatto.NatChem.2014, M.T.Huynh.ACSCS.2017, A.R.Offenbacher.JPCB.2010}}
specific pathways taken by electrons and protons, can lead to step-wise (consecutive) or concerted type reactions (CEPT). 
Unravelling the detailed mechanistic aspects of the PCET process  is fundamental for the design of artificial solar energy utilization systems, for example, dye sensitized photo-electrochemical cell (DS-PEC) and  many other bio-mimetic systems which have been developed for solar energy utilization and hydrogen reduction.{\cite{C.J.Gagliardi.CCR.2010, R.Becker.SciAdv.2016, A.Yamaguchi.NatCom.2014, J.F.Allen.Cell.2002, S.Papa.BBAB.2006}}. 

Various approximated quantum dynamical theories, mostly based on the determination of reaction rate constants, have been derived for isolated systems on the basis of the golden rule expression, linear response theory, and on Marcus's theory of electron-transfer (ET) processes. {\cite{R.I.Cukier.JPC.1996, Z.K.Goldsmith.FD.2019}}
Their applications have been extended to condensed phase systems by further assuming a perturbative system-bath interaction and a classical treatment of an environment representing, for example, solvent.{\cite{J.M.Mayer.JPCL.2011, A.V.Soudackov.FD.2016, A.V.Soudackov.JCP.2000, A.V.Soudackov.JCP.2015, R.I.Cukier.JPC.1994, R.I.Cukier.JPC.1996, A.Hazra.JCPB.2010}
Rate constants for several PCET systems in thermal equilibrium conditions have also been computed with the aids of molecular dynamics simulations and quantum chemistry calculations.{\cite{J.S.Kretchmer.IC.2016, M.N.Kobrak.JPCB.2001, J.Grimminger.CP.2007, B.Auer.JACS.2011,P.Andrea.JCP.2020}}

Yet, the sole computation of reaction rates does not provide enough information to fully disentangle  different ET and PT pathways and can hide important information about  the role of the environment.
Ultrafast nonlinear spectroscopy can be a powerful tool for unravelling the mechanistic aspects of PPCET reactions and of photosynthesis in general. {\cite{M.Cho.2019}} 
For example, infrared (IR) transient absorption spectroscopy (TAS)  has been applied to excited-state proton transfer and chemical bond cleavage, and can provide a versatile tool to determine the relaxation mechanism after an initial photoexcitation. {\cite{M.Pfeiffer.LC.1999, D.G.Hogle.JPCB.2018}
The results of luminescence TAS have indicated that the quantum effect of donor-acceptor (D-A) vibrations on PPCET is important for a full quantum treatment of the total reaction system. {\cite{M.K.Petermann.JACS.2012, Y.Giret.JCP.2020, R.Zheng.CP.2011, K.Song.JCP.2017, P.F.Barbara.Sci.1992}}

These spectroscopic techniques have also been extended to multi-dimensional cases. 
Two-dimensional (2D) vibrational spectroscopy (2DVS){\cite{Y.Tanimuka.JCP.1993, S.Mukamel.1995, P.Hamm.2011}} and 2D electronic spectroscopy (2DES) have been applied to condense phase transition and succeeded in investigating the electronic excitation dynamics and a structural change of molecules.{\cite{J.D.Gaynor.OL.2016,T.A.A.Oliver.RSOS.2017, Z.W.Fox.JPCL.2020, S.Mukamel.ARPC.2000, H.Dong.JCP.2015}}
Their combination, 2D electronic-vibrational spectroscopy (2DEVS), has also been successfully applied to the photo-isomerization reactions, mental-to-ligands transitions, conical intersection wavepacket dynamics, and ultrafast excitonic photosynthetic energy transfer reactions. {\cite{E.C.Wu.PCCP.2019,T.Ikeda.CP.2018, N.H.C.Lewis.JPCL.2016,T.L.Courtney.JCP.2015, E.C.Wu.FD.2019}}
By utilizing the UV-vis and IR pulses, we are now able to measure the coupling strength and coherence between the electronic and vibrational transitions as the off-diagonal peaks of 2D spectroscopy. {\cite{M.Cho.2019, S.Mukamel.1995}}
These features are useful for the investigation of PPCET reaction dynamics. 

In this paper, we present a model of a PPCET reaction and provide a detailed analysis of its dynamics by computing the signals of TAS and a newly defined 2D resonant electronic-vibrational spectroscopy (2DREVS). 
The 2DREVS is an extension of 2DEV for a strong resonant reaction system, and is useful for investigating the dynamics of PPCET reaction, as described below.
We describe the coupled proton-electron dynamics using two-dimensional potential energy surfaces (PESs), and complex system-bath interactions to simulate a system in realistic conditions.
We employ the numerically ``exact'' hierarchical equations of motion (HEOM) approach to study the reduced system dynamics under non-perturbative and non-Markovian system-bath interactions at finite temperature.
{\cite{Y.Tanimura.JPSJ.1989, Y.Tanimura.PRA.1990, A.Ishizaki.JPSJ.2005, Y.Tanimura.JPSJ.2006, Y.Tanimura.JCP.2014, Y.Tanimura.JCP.2015, Y.Tanimura.JCP.2020}}
The paper is organized as follows.
In Sec. {\ref{sec.model}}, we derive a system-bath model for a prototypical PPCET process and introduce the HEOM approach for  numerical simulation. 
The theory of nonlinear response functions is also briefly sketched in this section.
In Sec. {\ref{sec.result}}, we present the calculated TAS and 2DREVS results and analyze their profiles.

\section{Theory}
\label{sec.model}

\subsection{Model Hamiltonian}
\label{sec.system}

The system considered in the present work is depicted in Fig. \ref{fig:pcetsys}.
In the ground electronic state the proton is localized at bond distance from donor $D$, and the $D$-$H$ moiety is hydrogen bonded to the acceptor $A$.
\begin{figure}[ht]
\centering
\includegraphics[width=0.6\textwidth]{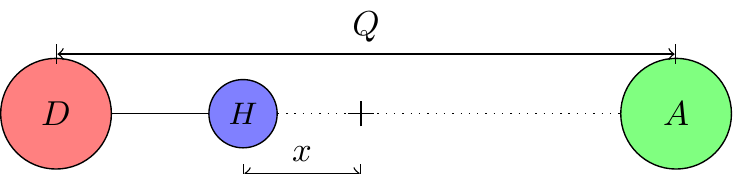}
\caption{\label{fig:pcetsys}
Model PPCET system with a hydrogen bridge. 
Here, $D (A)$ is the donor (acceptor), $H$ is the transferring proton, $x$ is the proton coordinate describing its distance from the center of the $D$ and $A$ units, and $Q$ is the distance between the heavy atoms connected by the hydrogen bond. 
}
\end{figure}
The $x$ coordinate describes the position of the proton between $D$ and $A$, while $Q$ is the distance between the heavy atoms which is also referred to as the reaction promoting mode. 
We wish to describe the dynamics of the system resulting from the photo-excitation of  $D$, which is followed by a coupled transfer of an electron and a proton to the $A$ moiety. 
As a result of the process an hydrogen atom is transferred from $D$-$H$ to $A$, {\textit{i.e.}} $A$ is reduced to $A$-$H$ and $D$-$H$ is oxidized $D$.

In order to model the coupled PT and ET processes we consider an electronic active space comprising the ground state $|\phi_g\rangle$ of the system, the localized excited state $|\phi_{LE}\rangle$, in which only the moiety $D$ is in the first excited electronic state, while $A$ is in the ground electronic state, and the charge-transfer state $|\phi_{CT}\rangle$, in which $D$ has transferred an electron to $A$. 
The diabatic representation of system is shown in Fig. {\ref{fig.potential_surface}}.
The motion along the $x$ and $Q$ coordinates is described employing realistic two-dimensional potential energy surfaces. 
Furthermore, we assume that the system interacts with a condensed phase environment which can be either a solvent or a protein scaffold. 
The overall Hamiltonian can therefore be expressed as
\begin{align}
\hat{H} = \sum_{i} \hat{H}_i (\hat{x}, \hat{Q},\{q_a\}) \big|i\rangle \langle i\big| 
+ \sum_{i\neq j} \Delta_{ij} \big|i\rangle \langle j\big|  + \hat{H}_B,
\label{eq.H_total}
\end{align}
where $\hat{H}_i (\hat{x}, \hat{Q},\{q_{a}\})$ is the Hamiltonian for the electronic states $i=g, CT$, and $LE$, $\Delta_{ij}$ are the electronic couplings among different electronic states, and $ \hat{H}_B$ is the Hamiltonian of the thermal bath which is modeled as a collection of harmonic oscillators
\begin{align}
\hat{H}_{B} = \sum_{a} \left( \frac{\hat{p}_a^2}{2m_a} + \frac{1}{2} m_{a} \omega_{a}^2 \hat{q}_{a}^2  \right) ,
\end{align}
where $\hat{p}_{a}$, $\hat{q}_{a}$, $m_{a}$ and $\omega_{a}$ are the momentum, position, mass and frequency of ${a}^{th}$ bath oscillator, respectively.

\begin{figure}[h]
\centering
\includegraphics[width=0.8\textwidth]{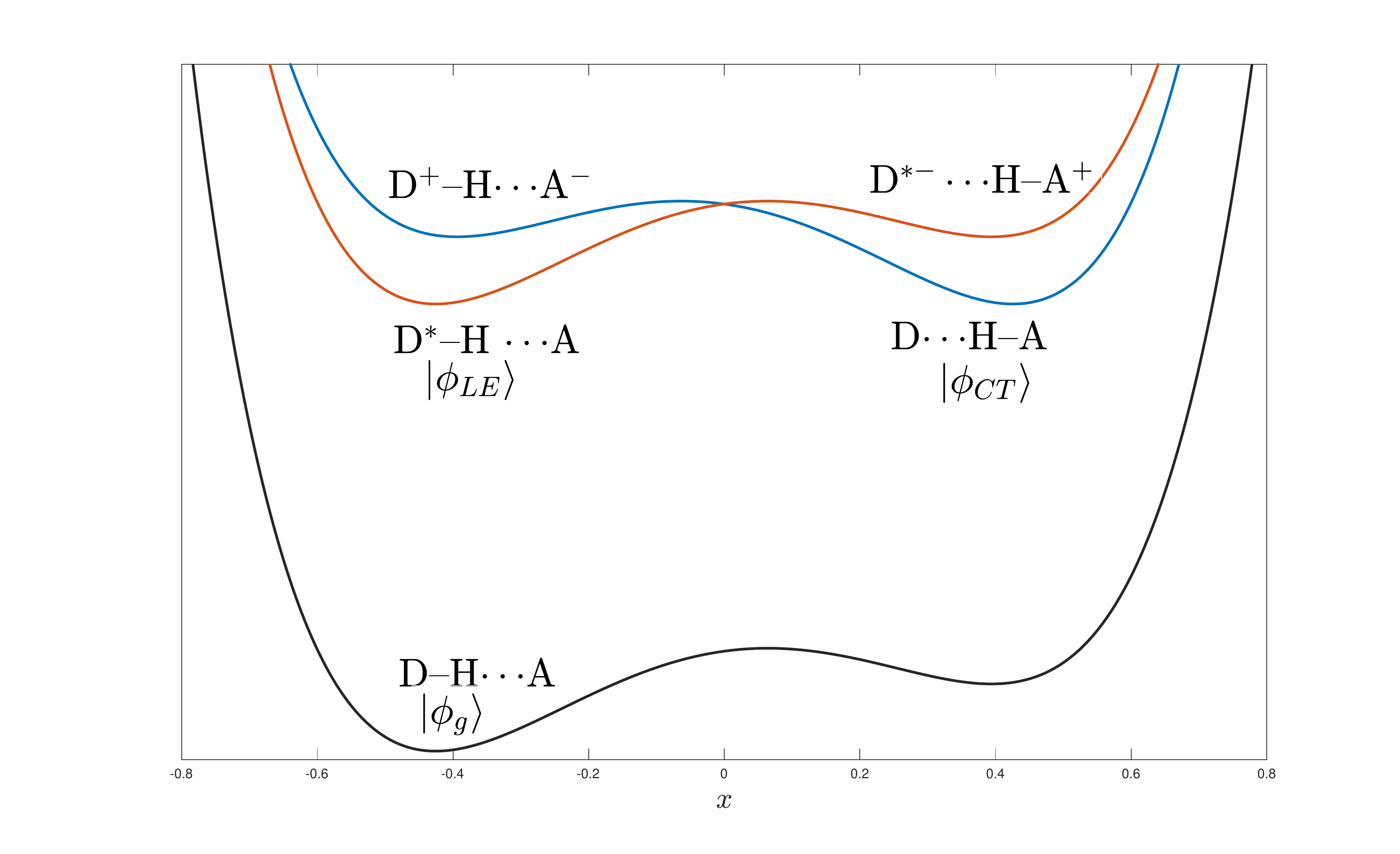}
\caption{
Diabatic representation of the PES for the reduced system. 
The black curve represent the ground state ${\left|\phi_{g}\right\rangle}$.
The red, and blue curves represent the local excited state ${\left|\phi_{LE}\right\rangle}$, and charge transfer state ${\left|\phi_{CT}\right\rangle}$, respectively.
}
\label{fig.potential_surface}     
\end{figure}

The operators $\hat{H}_i$ can be explicitly written in the form
\begin{align}
\hat{H}_i(\hat{x},\hat{Q},
\{q_{a}\}) =  \frac{\hat{p}_x^2}{2m_x}+\frac{\hat{p}_Q^2}{2m_Q} +U_i(\hat{x}, \hat{Q},\{q_{a}\}) + \varepsilon_i.
\end{align}
Here,  $\hat{x}$, $\hat{p}_x$,  $m_x$, and $\hat{Q}$, $\hat{p}_Q$ $m_Q$ are the coordinate, momentum and mass of the proton and of the D-A vibration, respectively, and $\varepsilon_i$ is the the energy of the electronic state for $i$.   
The potential $U_i(\hat{x}, \hat{Q},\{q_{a}\})$ describes the variation of the electronic energy as a function of the coordinates $x,Q$ and, furthermore, explicitly includes the interaction between these coordinates and the bath modes $\{q_{a}\}$. 
Following previous work, {\cite{J.Zhang.JCP.2020}} we use an asymmetric double well Morse potential for the proton mode, and a harmonic potential for the D-A mode 
\begin{align}
U_i(\hat{x}, \hat{Q}, \{q_{a}\} ) &= D_{i}^{l} \left[ 1 - e^{ -\alpha ( \hat{Q}/2 + \hat{x} - {x_e(\{q_{a}\})})} \right]^2  
+ D_{i}^{r} \left[ 1 - e^{ -\alpha ( \hat{Q}/2 - \hat {x} - x_e(\{q_{a}\}))} \right]^2  \nonumber \\
& + \frac{1}{2} D_k {(\hat Q - {Q_e})}^2,
\label{eq.U_i}
\end{align}
where $D_{i}^{l}$ and $D_{i}^{r}$ are the dissociation energy of donor (left well) and acceptor (right well), ${x_e}$ and ${Q_e}$ are the equilibrium distance of the proton and D-A vibrations, $\alpha$ represents the curvature of the Morse potentials, and  $D_k$ is the force constant of the D-A vibration. 
In our model the role of bath modes $\{q_{a}\}$ is to dynamically perturb the equilibrium position of the proton via a linear interaction, that is
\begin{align}
{x_e}(\{q_{a}\}) = {x_0} -  \sum_{a} g_{a} \hat{q}_{a} , 
\label{eq.interaction_origin}
\end{align}
where ${x_0}$ is the equilibrium distance without the heat bath, and $\{g_{a}\}$ are coupling strength parameters. 
Finally, we simplify this potential by expanding Eq. {\eqref{eq.U_i}} in terms of the collective coordinate $\hat{X} = \sum_{a} g_{a} q_{a}$ up to the first-order, which is similar to the reaction surface approach. {\cite{D.P.Tew.JCP.2006}}
The system PES and the resulting exponential-linear (EL) system-bath interaction are then expressed as
\begin{align}
{U_i}(\hat{x}, \hat{Q},X) =U_i^{0}(\hat x, \hat Q) + \hat{V}_{i}(\hat{x}, \hat{Q}) \hat{X}, 
\end{align}
where
\begin{align}
U_i^{0}(\hat{x}, \hat{Q}) &= D_{i}^{l} \left[ 1 - e^{ -\alpha ( \hat{Q}/2 + \hat{x} - x_0)} \right]^2  
+ D_{i}^{r} \left[ 1 - e^{ -\alpha (\hat{Q}/2 - \hat{x} - x_0)} \right]^2  \nonumber \\
&+ \frac{1}{2} D_K {(\hat{Q} - Q_0)}^2.
\end{align}
The operator $V_i$ depends solely on system variables and can be explicitly written as{\cite{J.Zhang.JCP.2020}}
\begin{align}
\hat{V}_{i}(\hat{x}, \hat{Q}) &= 2\alpha D_{i}^{l} \left[ 1 - e^{ -\alpha ( \hat{Q}/2 + \hat{x} - x_0)} \right] 
e^{-\alpha(\hat{Q}/2 + \hat{x}-x_0)}  \\
&+ 2\alpha D_{i}^{r} \left[ 1 - e^{ -\alpha ( \hat{Q}/2 - \hat{x} - x_0)} \right]
e^{-\alpha(\hat{Q}/2 -\hat{x}-x_0)}.
\label{eq.V_ix}
\end{align}
The structure of this rather complex form of system operator can be easily understood once we expand it in terms of $\hat{x}$ and $\hat{Q}$ as
\begin{equation}
\hat{V}_{i}(\hat{x}, \hat{Q}) = V_{i}^{(0)} + V_{i,x}^{(1)} \hat{x}+ V_{i,Q}^{(1)}\hat{Q}++ V_{i,x}^{(2)} \, \hat{x}^2+ V_{i,xQ}^{(2)} \, \hat{x} \hat{Q}+...,
\end{equation}  
where the $V_{i}^{(0)}$, $V_{i,x}^{(1)}$, etc. are  constants whose analytical expressions are given in Appendix {\ref{sec.appendix.01}}. 
 Hence it is clear that the coupling of Eq. {\eqref{eq.interaction_origin}} introduces linear interactions with the electronic subsystem via the constant $V_{i}^{(0)}$, and with the nuclear coordinates $\hat{x}$ and $\hat{Q}$ via $V_{i,x}^{(1)}$ and $V_{i,Q}^{(1)}$. 
As discussed in Ref. {\onlinecite{Y.Tanimura.JPSJ.2006}},  the linear-linear (LL) interaction, such as $ V_{i,x}^{(1)} \hat{x} q_{a}$ contributes mainly to energy relaxation, while the square-linear (SL) system-bath interaction, such as $ V_{i,x}^{(2)} \, \hat{x}^2 q_{a} $ leads to vibrational dephasing in the slow modulation case, due to the frequency fluctuation of the system vibrations.
Finally we note that the LL contribution in the proton mode vanishes for symmetric double well potential, i.e. $D_{i}^l=D_{i}^r$. {\cite{J.Zhang.JCP.2020}}

For simplicity, we further assume that all of the electronic states are coupled to the same heat bath. 
Then, Eq. {\eqref{eq.H_total}} can be rewritten as
\begin{align}
\hat{H}_{\rm tot} &= \sum_i \hat{H}_{i}^{0}{\big | i \big\rangle \big\langle i\big|}
+ \sum_{i\neq j}\Delta_{ij}{\big | i \big\rangle \big\langle j\big|} 
 + \hat{H}_{B+I}, 
\label{eq.H_total_alter}
\end{align}
where
\begin{align}
\hat{H}_i^{0} =  \frac{\hat p_x^2}{2m_x}+\frac{\hat p_Q^2}{2 m_Q} +U_i^{0}(\hat x, \hat Q)
\end{align}
and 
\begin{align}
  \hat{H}_{B+I} &=\sum _{a}\left\{\frac{\hat{p}_{a}}{2 m_a}+\frac{m_{a} \omega_{a}^{2}}{2}\left[\hat{q}_{a}-\frac{g_a
\left( \sum_i {\big | i \big\rangle \big\langle i\big|} \hat{V}_{i}\right)}{m_{a}} \omega_{a}^{2}\right]^{2}\right\}.
\end{align}
We also include the counter-term in the definition of $\hat{{H}}_{B+I}$ in order to maintain the translational symmetry of the system. {\cite{Y.Tanimura.JPSJ.2006, Y.Tanimura.JCP.2014}}

\subsection{Hierarchical Equations of Motion Approach}
\label{sec.heom}

Next, we briefly introduce the hierarchical equations of motion (HEOM) approach, which is employed to investigate quantum dynamics of the PCET system in a numerically rigorous way.{\cite{Y.Tanimura.JPSJ.2006, Y.Tanimura.JCP.2020}} 
We can also employ the multistate quantum hierarchical Fokker-Planck equations (MQHFPE), which has been applied to both optical and nonadiabatic transition problems described by complex PESs}. {\cite{Y.Tanimura.JCP.1997, T.Ikeda.JCP.2017, T.Ikeda.JCP.2019}} 
However, here we choose the regular HEOM in the energy eigenstate representation for both electronic and vibrational modes.
This is because the proton motion is well confined in the PESs, and the computational cost for using MQHFPE is much higher that regular HEOM. 

The heat bath is described by the spectral distribution function (SDF) 
\begin{align}
J(\omega) = \pi \sum_a \frac{g_a^2}{2 m_{{a}} \omega_a} \delta(\omega - \omega_a)
\end{align}
and the inverse temperature, $\beta \hbar = 1/ k_B T$, where $k_B$ is the Boltzmann constant. 
The overall noise effect on the system is characterized by the correlation function
\begin{align}
C(t) = \hbar \int_0^{\infty}  {\rm{d}}{\omega} J(\omega) \left[ {\coth} \left( \frac{\beta \hbar \omega}{2} \right) {\cos}(\omega t) - i {\sin}(\omega t) \right],
\label{eq.correlation_function}
\end{align}
where the notation $\langle \ldots \rangle_B$ represents the thermal average taken with the canonical distribution of the bath. 
In this paper, we use a Drude formed SDF,
\begin{align}
J(\omega) = \frac{\zeta}{2\pi} \frac{\omega \gamma^2}{\gamma^2 + \omega^2},
\end{align}
where $\zeta$ represents the coupling strength, and $\gamma$ is the reciprocal of the noise correlation time, representing the width of the spectral distribution.
Then, Eq. {\eqref{eq.correlation_function}} can be expressed in terms of a combination of linear exponential functions and of the $\delta(t)$ function, as 
\begin{align}
C(t) = \sum_{k=0}^{K} (c_k^{\prime} + i c_k^{\prime \prime}) \gamma_{k} e^{- \gamma_{k} t} 
+ 2c_{\delta} \cdot \delta (t), 
\label{eq.correlation_function_decompose}
\end{align}
where $c_{k}^{\prime}$, $c_{k}^{\prime\prime}$, $\gamma_{k}$ and $c_{\delta}$ are  constants determined by the chosen decomposition method.
Here we employ the Pad{\'e} decomposition method {\cite{J.Hu.JCP.2011, J.J.Ding.JCP.2012}} which is known to enhance the efficiency of numerical calculations. 
By introducing the auxiliary density operators (ADO) $\hat{\rho}_{\vec{n}}$, the HEOM can be derived as {\cite{Y.Tanimura.JPSJ.1989, Y.Tanimura.PRA.1990, A.Ishizaki.JPSJ.2005, Y.Tanimura.JPSJ.2006, Y.Tanimura.JCP.2014, Y.Tanimura.JCP.2015, Y.Tanimura.JCP.2020}}
\begin{align}
\frac{\partial}{\partial t} \hat{\rho}_{\vec{n}}(t)
= &\, - \left[ \frac{i}{\hbar} \mathcal{\hat L}_S 
+ \sum_{k} n_{k} \gamma_{k}+ c_{\delta} \hat{\Phi}^2 \right]  \hat{\rho}_{\vec{n}}(t) \nonumber \\
&\, - \sum_{k} \hat{\Phi} \hat{\rho}_{\vec{n} + \vec{e}_{k}}(t)
 - \sum_{k} n_{k} \hat{\Theta}_{k}  \hat{\rho}_{\vec{n} - \vec{e}_{k}}(t),
\label{eq.HEOM}
\end{align}
where the superoperators are defined as {${\mathcal {\hat L}}_S {\hat A} \equiv [ { {\hat H}_S}, { {\hat A}} ]$ and  $\hat{\Theta}_{k} \equiv c_{k}^{\prime} \hat{\Phi} - c_{k}^{\prime \prime} \hat{\Psi}$ with
\begin{align}
{\hat{\Phi}} {\hat A} \equiv \frac{i}{\hbar} \left[ \sum_i {\big | i \big\rangle \big\langle i\big|} \hat{V}_{i}, { {\hat A}} \right],~~~ 
{\hat \Psi} {\hat A} \equiv  \frac{1}{\hbar} \left\{ \sum_i {\big | i \big\rangle \big\langle i\big|} \hat{V}_{i}, { {\hat A}} \right\},
\end{align}
for any physical operator $\hat A$}. 
The components of multi-index vector  $\vec{n} = (..., n_{k}, ...)$ are all non-negative integers, and $\vec{e}_{k}$ is the $k^{th}$ unit vector. 
In HEOM formalism, only the first element, $\vec{n} = (0,...,0)$, has a physical meaning, corresponding to the reduced density operator of system. 
The others are served as the treatment of non-perturbative and non-Markovian heat bath effect. {\cite{Y.Tanimura.JPSJ.2006, Y.Tanimura.JCP.2020}}
Although Eq. {\eqref{eq.HEOM}} consists of infinite equations, we can truncate it at a properly chosen large $N$ value, for $N = \sum_{k} n_{k}$. {\cite{A.Ishizaki.JPSJ.2005}}
In order to reduce the computational cost for the time integration, we rescale the ADOs as $\hat{\rho}_{\vec{n}}= \hat{\rho}_{\vec{n}}/{\prod_{u,k} \sqrt{n_{u,k}!}}$.
Then, Eqs. {\eqref{eq.HEOM}} are rewritten as {\cite{Q.Shi.JCP.2009, T.Ikeda.JCTC.2019}}
\begin{align}
\frac{\partial}{\partial t} \hat{\rho}_{\vec{n}}(t)
= &\, - \left[ \frac{i}{\hbar} \mathcal{\hat L}_S 
+ \sum_{k} n_{k} \gamma_{k}+ c_{\delta} \hat{\Phi}^2 \right]
\hat{\rho}_{\vec{n}}(t) \nonumber \\
&\, - \sum_{k} \sqrt{n_{k}+1} \hat{\Phi} \hat{\rho}_{\vec{n} + \vec{e}_{k}}(t) 
- \sum_{k} \sqrt{n_{k}} \hat{\Theta}_{k} \hat{\rho}_{\vec{n} - \vec{e}_{k}}(t).
\label{eq.HEOM_filter}
\end{align}

\subsection{Projection operators for PT and ET states}

In order to analyze the PCET process, next we introduce a set of projection operators defined as 
\begin{align}
\hat{\mathcal{\theta}}_{i}^l = \big|i\rangle\langle i\big| \hat{h}(-x),~~~
\hat{\mathcal{\theta}}_{i}^r=  \big|i\rangle\langle i\big|\hat{h}(x), 
\end{align}
where $i=CT$ and $LE$, $\hat{h}(x)$ is the Heaviside step function for {the} proton coordinate, and the symbols $l$ and $r$ represent the proton localized in the left (donor) and right (acceptor) well, respectively. 
The corresponding population of the superposition is $P_i^{\alpha} (t)= \rm{Tr}\left\{\hat{\mathcal{\theta}}_i^{\alpha} \hat{\rho}(t)\right\}$ for $\alpha = l$ or $r$.
The populations of ${\left|\phi_{LE}\right\rangle}$ and ${\left|\phi_{CT}\right\rangle}$ is then separated as $P_i (t) = P_i^l (t)+P_i^r (t)$, whereas that in the left and right well is expressed as $P^{\alpha} (t)= P_{LE}^{\alpha} (t)+P_{CT}^{\alpha}(t)$.  

As shown in Fig. {\ref{fig.potential_surface}}, the superposition ${\left|\phi_{LE}^{l}\right\rangle}$ represents the configuration $D^*-H\cdots A$, and ${\left|\phi_{LE}^{r}\right\rangle}$ represents ${\rm{D^*}}^{-}\cdots H - {\rm{A}}^{+}$. 
Similarly, ${\left|\phi_{CT}^{l}\right\rangle}$ represents $D^{+}-H\cdots A^{-}$, and ${\left|\phi_{CT}^{r}\right\rangle}$ represents ${\rm{D}}\cdots H - {\rm{A}}$.
Thus, the pure PT process corresponds to the transitions ${\left|\phi_{LE}^{l}\right\rangle} \leftrightarrow {\left|\phi_{LE}^{r}\right\rangle}$ and ${\left|\phi_{CT}^{l}\right\rangle} \leftrightarrow {\left|\phi_{CT}^{r}\right\rangle}$. 
The pure ET process corresponds to the transitions ${\left|\phi_{LE}^{l}\right\rangle} \leftrightarrow{\left|\phi_{CT}^{l}\right\rangle}$ and ${\left|\phi_{LE}^{r}\right\rangle} \leftrightarrow {\left|\phi_{CT}^{r}\right\rangle}$. 
The CEPT process corresponds to the transitions ${\left|\phi_{LE}^{l}\right\rangle} \leftrightarrow {\left|\phi_{CT}^{r}\right\rangle}$ and ${\left|\phi_{LE}^{r}\right\rangle} \leftrightarrow {\left|\phi_{CT}^{l}\right\rangle}$.

\subsection{Nonlinear Response Function}
\label{sec.resp}

The nonlinear response functions can be calculated within the framework of the HEOM formalism. {\cite{Y.Tanimura.JPSJ.2006, Y.Tanimura.JCP.2020}} 
The third-order {optical} response function can be expressed as
\begin{align}
R^{(3)} (t_3,t_2,t_1) = {\left( \frac{i}{\hbar} \right)}^{3} {\rm{Tr}}  \left\{ \hat{\mu}_4 \mathcal{G}(t_3) 
\hat{\mu}_3^{\times} \mathcal{G} (t_2) \hat{\mu}_2^{\times} \mathcal{G}(t_1) \hat{\mu}_1^{\times} {\hat{\rho}^{eq}} \right\},
\label{eq.resp_3rd_time}
\end{align}
where $\hat{\mu}_{k}$ is the dipole operator of the $k$-th laser interaction, ${\mathcal{G}}(t)$ is the Green's function of the total Hamiltonian without laser interactions, and ${\hat{\rho}^{eq}}$ is the initial state density operator. 
In the HEOM approach, the density matrix is replaced by a reduced one, and ${\mathcal{G}}(t)$ is evaluated from Eq. {\eqref{eq.HEOM}} (or Eq.\eqref{eq.HEOM_filter}). \cite{Y.Tanimura.JPSJ.2006}
The operator $\hat{\mu}_{k}^{\times}$ is the commutator of the dipole operator $\hat{\mu}_i$.
The right-hand side of Eq. {\eqref{eq.resp_3rd_time}} can be evaluated as follows: 
The system is first in the initial equilibrium state ${\hat{\rho}^{eq}}$, and is excited by the first interaction $\hat{\mu}_1^{\times}$ at $t=0$. 
The time evolution is computed by numerically integrating Eq. \eqref{eq.HEOM} up to a chosen time $t_1$.  
Then, the system is excited by the second and third interactions $\hat{\mu}_2^{\times}$ and $\hat{\mu}_3^{\times}$ in a similar way.
The final signal is computed by the expectation value of $\hat{\mu}_4$.
We compute $R^{(3)} (t_3,t_2,t_1)$ for a set of values of $t_1$, $t_2$, and $t_3$. 

Here we assume that the PES of $|\phi_{g}\rangle$ and $|\phi_{LE}\rangle$ have the same equilibrium positions, their energy difference is large and the population relaxation in the excited states is small.  
The direct excitation from ${\left|\phi_{g}\right\rangle}$ to ${\left|\phi_{CT}\right\rangle}$ is also prohibited. 
Thus, the initial state is described by the thermal equilibrium distribution of the $|\phi_{g}\rangle$ as ${\hat{\rho}^{eq}}={\hat{\rho}_{g}^{eq}}$.  
Assuming the PPCET reaction is initialized by a pair of impulsive pump pulses that excite the system from ${\left|\phi_{g}\right\rangle}$ to ${\left|\phi_{LE}\right\rangle}$, we set the initial conditions as ${\hat{\rho}}^{(2)}(0) =  -\hat{\mu}_1^{\times} \hat{\mu}_2^{\times}{\hat{\rho}_{g}^{eq}}/\hbar^2$ for further response function analysis. 
With the previous assumption, we can further set ${\hat{\rho}}^{(2)}(0) = {\hat{\rho}_{LE}^{eq}}$, where ${\hat{\rho}_{LE}^{eq}}$ is evaluated as the steady state solution of the HEOM for the $|\phi_{LE}\rangle$ state without non-adiabatic coupling with the $|\phi_{CT}\rangle$. 
Thus, our discussion in the following only considers the dynamics between ${\left|\phi_{LE}\right\rangle}$ and ${\left|\phi_{CT}\right\rangle}$.

The transient absorption response function can be evaluated from Eq. {\eqref{eq.resp_3rd_time}} by keeping $t_1 = 0$ as
\begin{align}
R^{\mathrm{TA}}(t ,t^{\prime})= \frac{i}{\hbar}  {\rm{Tr}}  \left\{ \hat{\mu}_4 \mathcal{G}(t) \hat{\mu}_3^{\times} \mathcal{G} (t^{\prime}) {\hat{\rho}}^{(2)}(0) \right\}.
\label{eq:ta_response}
\end{align}
Transient absorption spectrum (TAS) at different $t^{\prime}$ is evaluated as 
\begin{align}
I^{\mathrm{TA}}(\omega ,t^{\prime})&\equiv \omega \mathrm{Im}\int_{0}^{\infty } {\rm{d}}t e^{i\omega t }R^{\mathrm{TA}}(t,t^{\prime}),
\label{eq:ta-spectrum}
\end{align}
which also corresponds to linear absorption spectrum for non-equilibrium initial conditions. 
For the calculations of TAS, the dipole operators $\hat{\mu}_k$ for $k\ge3$ are assumed to be either electron part $\hat{\mu}_e$, or proton part $\hat{\mu}_p$, defined as
\begin{align}
\hat{\mu}_{e} &= {\left|\phi_{LE}\right\rangle} {\left\langle \phi_{CT} \right|}  
+ {\left|\phi_{CT}\right\rangle} {\left\langle \phi_{LE} \right|}, \nonumber\\
\hat{\mu}_{p} &= \hat{x} \cdot \left( {\left|\phi_{LE}\right\rangle} {\left\langle \phi_{LE} \right|} 
+  {\left|\phi_{CT}\right\rangle} {\left\langle \phi_{CT} \right|} \right).
\end{align}
Here, $\hat{\mu}_e$ is for spectroscopy of the electronic subsystem, and $\hat{\mu}_p$ is for spectroscopy of vibrational degrees of freedom, respectively.

The fifth-order transient 2D spectroscopy is defined in a similar way as
\begin{align}
R^{\mathrm{(5)}} (t_3,t_2,t_1) = {\left( \frac{i}{\hbar} \right)}^{3} {\rm{Tr}} 
\left\{ \hat{\mu}_6 \mathcal{G}(t_3) \hat{\mu}_5^{\times} {\mathcal{G}} (t_2) 
\hat{\mu}_4^{\times} \mathcal{G} (t_1) \hat{\mu}_3^{\times} {\hat{\rho}}^{(2)}(0)  \right\},
\label{eq.resp_5th_time}
\end{align}
where ${\hat{\rho}}^{(2)}(0) $ is the same as TAS. 
The transient 2D correlation spectroscopy are then evaluated as
\begin{align}
I^{\mathrm{Corr}}(\omega _{3},t_2,\omega _{1}) & = I^{\mathrm{(NR)}}(\omega _{3}, t_2^{\prime},\omega _{1}) + I^{\mathrm{(R)}}
(\omega _{3},t_2,\omega _{1}),
\label{eq.2D_correlation}
\end{align}
where the non-rephasing and rephrasing parts of the signal are expressed as
\begin{align} 
I^{\mathrm{NR}}(\omega_3, t_2, \omega _1) &= \mathrm{Im} \int _{0}^{\infty }{\rm{d}} t_3 \int_0^\infty
d t_1 e^{i \omega_3 t_3} e^{i \omega _1 t_1} R^{\mathrm{(5)}}(t_3,t_2,t_1), \\
I^{\mathrm{R}}(\omega_3, t_2, \omega _1) &= \mathrm{Im} \int _{0}^{\infty }{\rm{d}} t_3 \int_0^\infty
d t_1 e^{i \omega_3 t_3} e^{-i \omega _1 t_1} R^{\mathrm{(5)}}(t_3,t_2,t_1).
\end{align}
In a typical system measured  by 2DEVS, the frequency of electronic excitation is much higher than vibrational modes and the signals only have off-diagonal components.
However, in our model, the energy levels of electron and proton are similar, as shown in Figs. {\ref{fig.potential_surface}} and {\ref{fig.energy_level}}.  Thus, it may not easy to excite either electron or proton modes separately. Thus, here we assume the dipole operators to be the summation of both electron and proton, $\hat{\mu}_k = \hat{\mu}_e + \hat{\mu}_p $ for $k\ge3$.  The signals in this measurement then be a summation of 2DES, 2DEV, and 2DVS and is refereed to as 2D resonant electronic vibrational spectroscopy (2DREVS).

\begin{table}[h]
\caption{System parameters}
\centering
\begin{tabular}{c|c}
\hline
\hline
$\alpha$ & 2.0 $\rm{\AA}^{-1}$ \\
\hline
$x_0$ & 1.0 $\rm{\AA}$ \\
\hline
$Q_0$ & 3.0 $\rm{\AA}$ \\
\hline
$D_k$ & 303435 $\rm{cm}^{-1} \rm{\AA}^{-2}$ \\
\hline
$\Delta$ & 50 $\rm{cm}^{-1}$ \\
\hline
\hline
$D_{LE}^{l}$ & ~33715 $\rm{cm}^{-1}$  \\ 
\hline
$D_{LE}^{r}$ & ~31715 $\rm{cm}^{-1}$  \\ 
\hline
\hline
 $D_{CT}^{l}$ & ~31715 $\rm{cm}^{-1}$  \\ 
\hline
 $D_{CT}^{r}$ & ~33715 $\rm{cm}^{-1}$  \\ 
\hline
\hline
\end{tabular}
\label{table.parameters}
\end{table}

\begin{table}[h]
\caption{The lowest 10 energy eigenvalues of each electronic state as a unit of $\omega_0$. }
\centering
\begin{tabular}{c|c|c}
\hline
Eigen numbers $(m,n)$ & $\left|\phi_{LE}^{(m,n)}\right\rangle$ & $\left|\phi_{CT}^{(m,n)}\right\rangle$ \\[1.0ex]
\hline
\hline
(0, 0) & 0.00 & -0.02\\
\hline
(0, 1) & 0.82 & 0.81\\
\hline
(0, 2) & 1.63 & 1.64\\
\hline
(1, 0) & 1.94 & 1.95\\
\hline
(0, 3) & 2.44 & 2.47\\
\hline
(1, 1) & 2.78 & 2.74\\
\hline
(2, 0) & 3.06 & 3.06\\
\hline
(0, 4) & 3.26 & 3.31\\
\hline
(1, 2) & 3.64 & 3.55\\
\hline
(2, 1) & 3.94 & 3.97\\
\hline
\hline
\end{tabular}
\label{table.eigen_energy}
\end{table}

\section{Numerical Results}
\label{sec.result}

The system parameters chosen to simulate our PPCET model are listed in Table. {\ref{table.parameters}}, based on a typical PT system. {\cite{N.Sato.JCP.1988}}
The determination of the electronic couplings $\Delta$ is a critical point of any PPCET reaction, in that it provides the major contribution to the discrimination between adiabatic and non-adiabatic mechanisms. 
Here, we choose to study the system under moderate non-adiabatic conditions, and set $\Delta = 50~\mathrm{cm^{-1}}$, which is  close to previously reported studies. {\cite{S.Hammes-Schiffer.ChemRev.2010}}
The energy eigenstates of the system ${\big|\phi_{i}^{(m,n)}\big\rangle}$ are obtained by diagonalizing the matrix representation of the system Hamiltonian.  
The energy eigenvalues of the lowest several states are presented in Table. {\ref{table.eigen_energy}}, and a schematic view of is given in Fig. {\ref{fig.energy_level}}.
Here, the $m$ and $n$ represent the quantum numbers of the proton and D-A modes that are determined from the number of nodes along the $x$ and $Q$ directions. 
While, ${\big|\phi_{i}^{(m,n)}\big\rangle}$ with $m=0$ mainly correspond to the charge localized states ${\big|\phi_{LE}^{l}\big\rangle}$ and ${\big|\phi_{CT}^{r}\big\rangle}$, those with $m=1$ mainly corresponded to the intermediate transition states ${\left|\phi_{LE}^{r}\right\rangle}$ and ${\left|\phi_{CT}^{l}\right\rangle}$, respectively. 
The states for $m \geqq 2$ are strongly delocalized along the proton coordinate, and provide almost no contribution to the pure PT processes.
The numerical simulations of the HEOM were conducted using the energy eigenstates representation, and we employed the lowest 20-40 eigenstates for each electronic state based on the value of system-bath coupling strength $\zeta$.
The time integrals were carried out using the low-storage fourth-order Runge-Kutta (LSRK4) method.
The time step was chosen as $\delta t=0.01\omega_0^{-1}$, where $\omega_0$ is a characteristic frequency taken as the unit for all the other physical variables. 
Here, we choose $\omega_0=500{\rm{cm^{-1}}}$.
We also fixed the inverse correlation time as $\gamma =  0.5\omega_0$ and the bath temperature as $\beta\hbar\omega_0 = 2.4$ (300K).
The HEOM parameters required for a converged calculation were chosen as $N = 10$ and $K = 5$. 
In the following, we investigate the effects of the environment on the PCET mechanism as a function of $\zeta$ by studying both the population dynamics and the TAS and 2DREVS signals. 

\begin{figure}[t]
\centering
\includegraphics[width=0.6\textwidth]{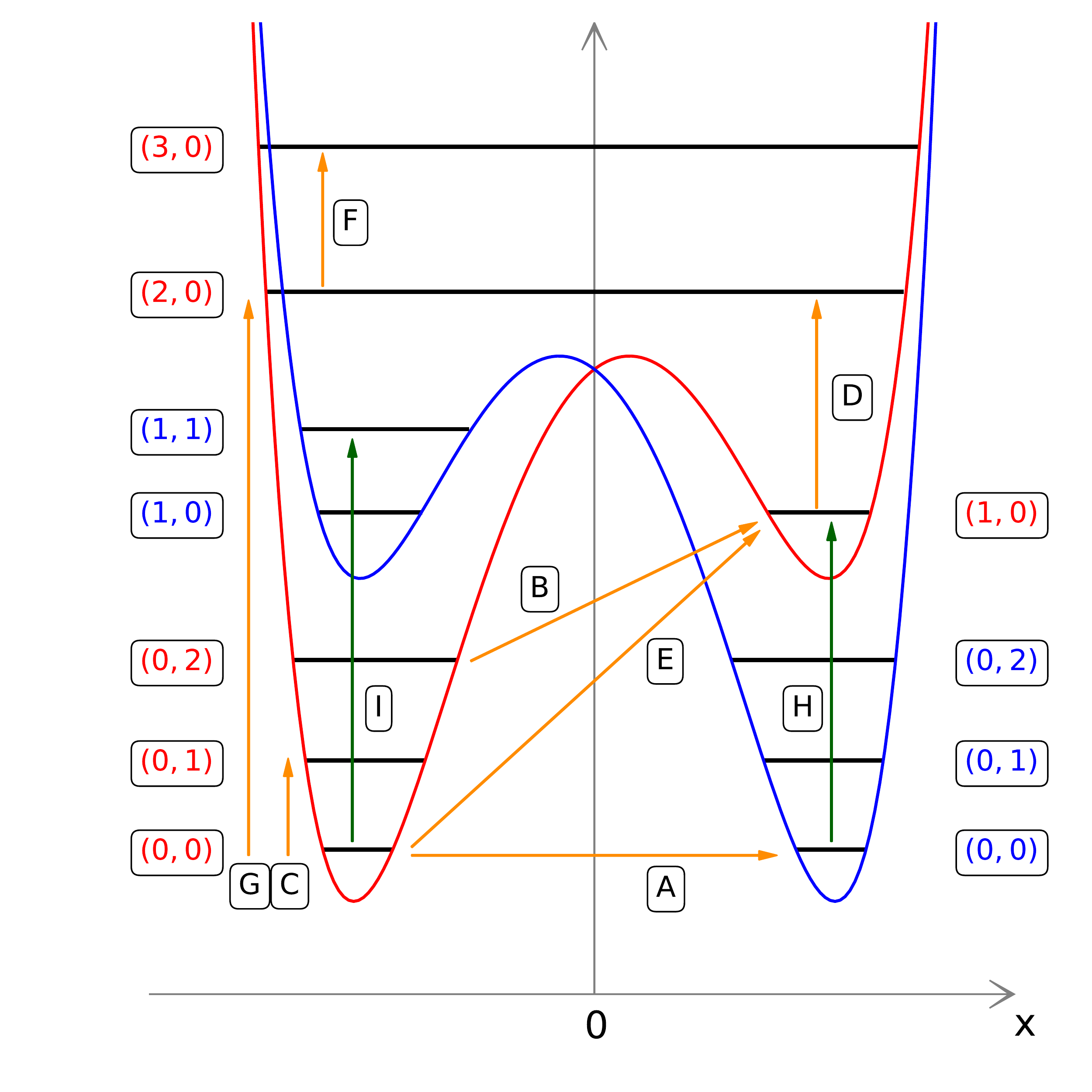}
\caption{A schematic view for ${\left|\phi_{LE}\right\rangle}$ (red curve) and ${\left|\phi_{CT}\right\rangle}$ (blue curve) in the diabatic representation along $\hat{x}$ at the minimum of $\hat{Q}$.  
The lowest several eigenstates for each PES are also plotted.  
The labeled orange and green arrows represent the corresponding proton and electron transitions appearing in nonlinear spectroscopy. 
See main text for the meaning of the labels.} 
\label{fig.energy_level}     
\end{figure}

\subsection{Population dynamics}
\label{sec.population}

\begin{figure}[b]
\centering
\includegraphics[width=\textwidth]{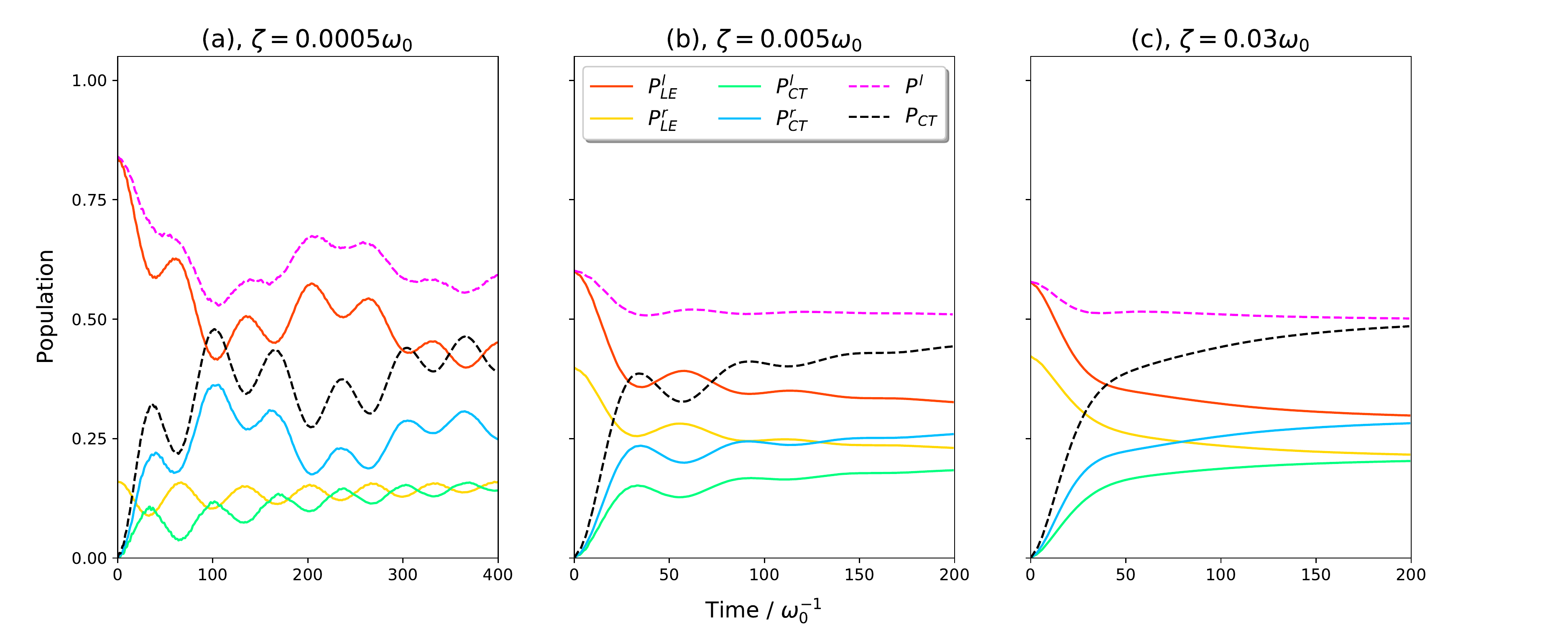}
\caption{The population dynamics that represents proton and electron localization. 
The red, yellow, green and blue curves represent $P_{LE}^l(t)$,  $P_{LE}^r(t)$,  $P_{CT}^l(t)$, and $P_{CT}^r(t)$, respectively, and the corresponding configurations of model system are illustrated in Fig. {\ref{fig.potential_surface}}.
The population $P^{l}(t)$ and $P_{CT}(t)$ are also presented as the dashed purple and black curves.} 
\label{fig.step_zeta_x}
\end{figure}

First, we illustrate the time evolution of the population states for various values of $\zeta$. 
The electron and proton transfer rates can be estimated from  $P_{CT}(t)$ and $P^{l}(t)$.
The calculated results are depicted in Fig. {\ref{fig.step_zeta_x}}, for {\hyperref[fig.step_zeta_x]{(a)}} a weak $(\zeta = 0.0005 \omega_0)$, {\hyperref[fig.step_zeta_x]{(b)}} moderate $(\zeta = 0.005 \omega_0)$, and {\hyperref[fig.step_zeta_x]{(c)}} strong $(\zeta = 0.03 \omega_0)$ coupling cases. 
Note that, as illustrated in our PT investigation, {\cite{J.Zhang.JCP.2020} the effective coupling strength on the present exponential-linear system-bath coupling model is different from the conventional linear-linear coupling models.
The strength of the coupling parameter is determined on the basis of the relaxation dynamics of the populations and spectral line shape of TAS, as we will show below.  

In the weak coupling case, Fig. {\hyperref[fig.step_zeta_x]{\ref{fig.step_zeta_x}(a)}}, coherent recursive oscillations of state populations are observed. 
These oscillations do not affect  the equilibrium distribution and do not contribute to the population transfer rates. 
Although the contribution is minor, the population exchange between $P_{CT}(t)$ and $P^{l}(t)$ suggests the presence of a charge transfer process.
For the moderate and strong coupling cases in Fig. {\hyperref[fig.step_zeta_x]{\ref{fig.step_zeta_x}(b)}} and {\hyperref[fig.step_zeta_x]{(c)}}, the linear term of $\hat{V}_{i}$ causes the population relaxation suppressing the coherent oscillations.
In the $\hat{x}$ direction (proton mode), the nonlinear terms of $\hat{V}_{i}$ also lead to a decrease of the energy barrier so that proton transfer is promoted.
In the $\hat{Q}$ direction, the linear term of $\hat{V}_{i}$ leads to a decrease of proton distance from the heavy atoms and increase the PT efficiency. 
A constant term $\hat{V}_i(0,0)$ (see Eq. \eqref{eq:V00}) is also present, corresponding to the interaction between the electronic states and the heat bath.
As a result, for larger $\zeta$, both the electron and the proton are equally distributed in the two wells because of the symmetric PES.
Note that we cannot disentangle the contribution from the CEPT, ET, and PT processes only from the analysis of population dynamics because they are mixed in the population states.

\subsection{Transient absorption spectroscopy (TAS) }

\begin{figure}[t]
\centering
\includegraphics[width=\textwidth]{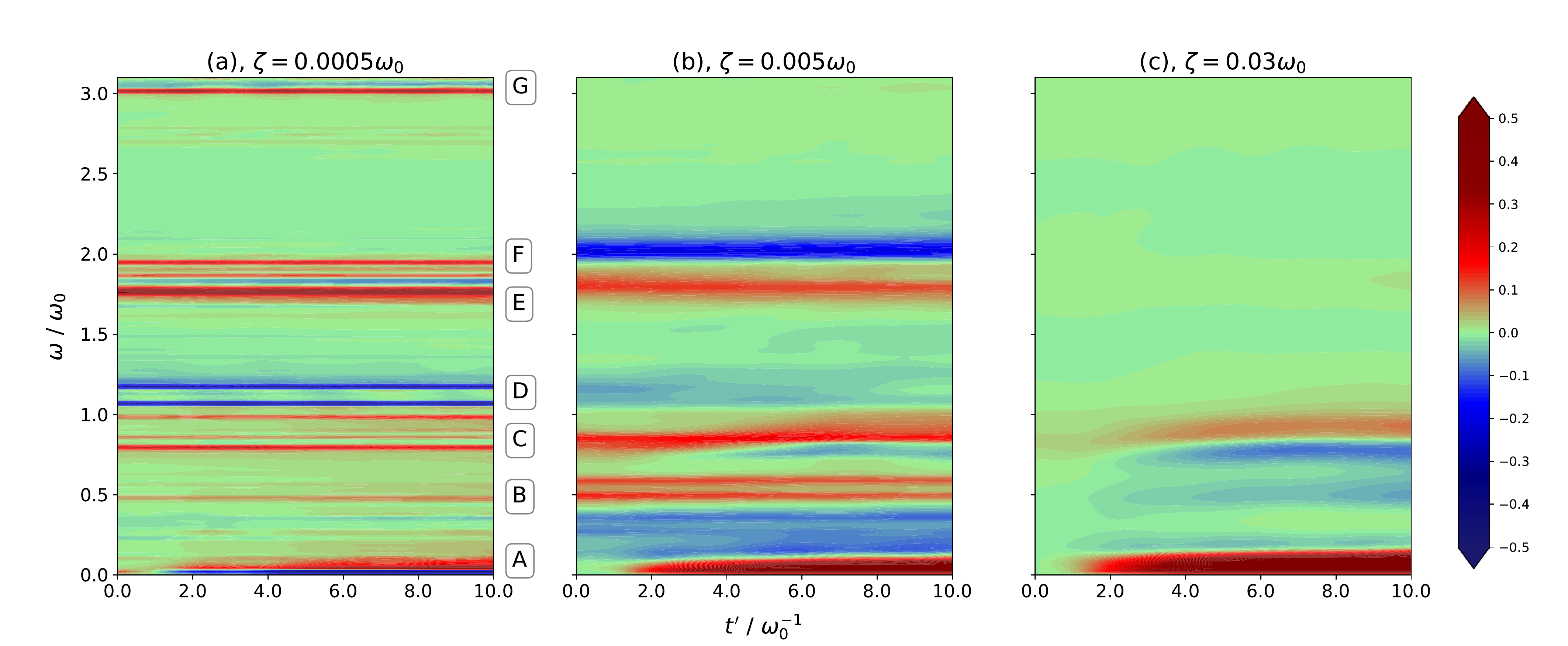}
\caption{The contour map of TAS calculated for the vibrational excitation  ( $\hat{\mu}=\hat{\mu}_p$) in the cases of the (a) weak, (b) moderate, and (c) strong system-bath interactions.
All the peaks are normalized with respect to maximum intensity of $t^{\prime}=10.0~\omega_0^{-1}$.
The contours are drawn from -0.5 to 0.5.
The red and blue areas represent the positive absorption and negative emission, respectively. } 
\label{fig.transient_proton_zeta_x}
\end{figure}

Next we present the results of TAS analysis.
Although TAS has the capability to analyze the populations in the ET and PT states separately following the position of the absorption peaks, this is not easy in the present case, because the excitation energies of the ET and PT processes are similar, and the abortion peaks are often overlapped. 
Hence, here we calculated TAS for the electronic and vibrational modes separately to help the analysis of 2DEVS.
In TAS, the charge transition rate can be evaluated from the intensity of corresponding transition peaks, while coherent oscillation appears as a $\delta$-function like peak. 
The characteristic time scale of various transitions can also be evaluated as a function of $t^{\prime}$.

In Fig. {\ref{fig.transient_proton_zeta_x}} we present TAS for the vibrational excitation of the proton mode obtained for a waiting time up to $t^{\prime} = 10.0 \omega_0^{-1}$ and by  setting $\hat{\mu}_3=\hat{\mu}_4=\hat{\mu}_p$.
In each figure, the negative and positive peaks represent the emission and absorption, respectively. 
Note that although the energy eigenvalues of the ${\big|\phi_{LE}^{(m,n)}\big\rangle}$ and ${\big|\phi_{CT}^{(m,n)}\big\rangle}$ in the diabatic representation are degenerate, those in the adiabatic representation are separated by the frequency $\Delta$ because of the diabatic coupling.

In the weak coupling case, Fig. {\hyperref[fig.transient_proton_zeta_x]{\ref{fig.transient_proton_zeta_x}(a)}}, the peak ``A'' $(0.05 \omega_0)$ predominantly arises from the CPET, ${\big|\phi_{i}^{(m,n)}\big\rangle} \rightarrow{\big|\phi_{j}^{(m,n)}\big\rangle}$. 
This transition always occurs due to the large overlap between two electronic potential surfaces.
The peak ``B'' $(0.5 \omega_0)$ and ``E'' $(2.7 \omega_0)$ arise from the pure PT with and without the participation of the D-A mode, where ``E'' represents ${\big|\phi_{i}^{(0,n)}\big\rangle} \rightarrow{\big|\phi_{i}^{(1,n)}\big\rangle}$, and ``B'' represents ${\big|\phi_{i}^{(0,n)}\big\rangle} \rightarrow{\big|\phi_{i}^{(1,n-2)}\big\rangle}$.
The peak ``C'' $(0.8 \omega_0)$ represents the excitation of the D-A mode, ${\big|\phi_{i}^{(m,n)}\big\rangle} \rightarrow{\big|\phi_{i}^{(m,n+1)}\big\rangle}$, which arises because the proton and the D-A mode are strongly coupled. 
The proton distribution varies as a function of the quantum number $n$ in the D-A mode, even when the quantum number of the proton mode $m$ is unchanged. 
The other three peaks represent the delocalization of the proton in the higher energy states ($m \leqq 2$), which do not contribute to either PT or CEPT.
The peaks ``D'' $(1.1 \omega_0)$, ``F'' $(1.9 \omega_0)$, and ``G'' $(3.0 \omega_0)$ represent ${\big|\phi_{i}^{(1,n)}\big\rangle} \rightarrow{\big|\phi_{i}^{(2,n)}\big\rangle}$, ${\big|\phi_{i}^{(2,n)}\big\rangle} \rightarrow{\big|\phi_{i}^{(3,n)}\big\rangle}$, and  ${\big|\phi_{i}^{(0,n)}\big\rangle} \rightarrow{\big|\phi_{i}^{(2,n)}\big\rangle}$, respectively.
Most of these peaks consist of several small peaks because of the participation of the D-A mode excited states $(n>0)$.
A schematic view of all the transitions is illustrated in Fig. {\ref{fig.energy_level}}.

We then analyze the effects of the system-bath coupling strength, $\zeta$, through the peak intensities as a function of $t^{\prime}$.
In the weak coupling case, Fig. {\hyperref[fig.transient_proton_zeta_x]{\ref{fig.transient_proton_zeta_x}(a)}}, most of the peaks are unchanged regardless of $t^{\prime}$ except for the peak ``A'', whose intensity changes sign near $t^{\prime} = 1.0$.
In the moderate and strong coupling cases, Fig. {\hyperref[fig.transient_proton_zeta_x]{\ref{fig.transient_proton_zeta_x}(b)}} and {\hyperref[fig.transient_proton_zeta_x]{\ref{fig.transient_proton_zeta_x}(c)}}, the peak intensity of ``A'' changes from almost 0 to a positive value in the initial time period.
This indicates that the CEPT process is promoted by the system-bath interaction, and occurs in a relatively short time period.
The promotion effect can be explained by the linear term $V_{i,Q}^{(1)}$ in the $\hat{Q}$ direction, which reduces the transfer distance and enhances the vibronic coupling.
The intensity of the peak ``C'' changes from positive to negative values near $t^{\prime} = 4.0$ in the case {\hyperref[fig.transient_proton_zeta_x]{\ref{fig.transient_proton_zeta_x}(b)}}, and $t^{\prime} = 2.0$ in the case {\hyperref[fig.transient_proton_zeta_x]{\ref{fig.transient_proton_zeta_x}(c)}}.
This implies that the characteristic time scale of the D-A excitation is larger than the CEPT.
The intensities of the PT peaks ``B'' and ``E'' are almost unchanged, which indicates that the pure PT plays a minor role in the present case.

\begin{figure}[h]
\centering
\includegraphics[width=\textwidth]{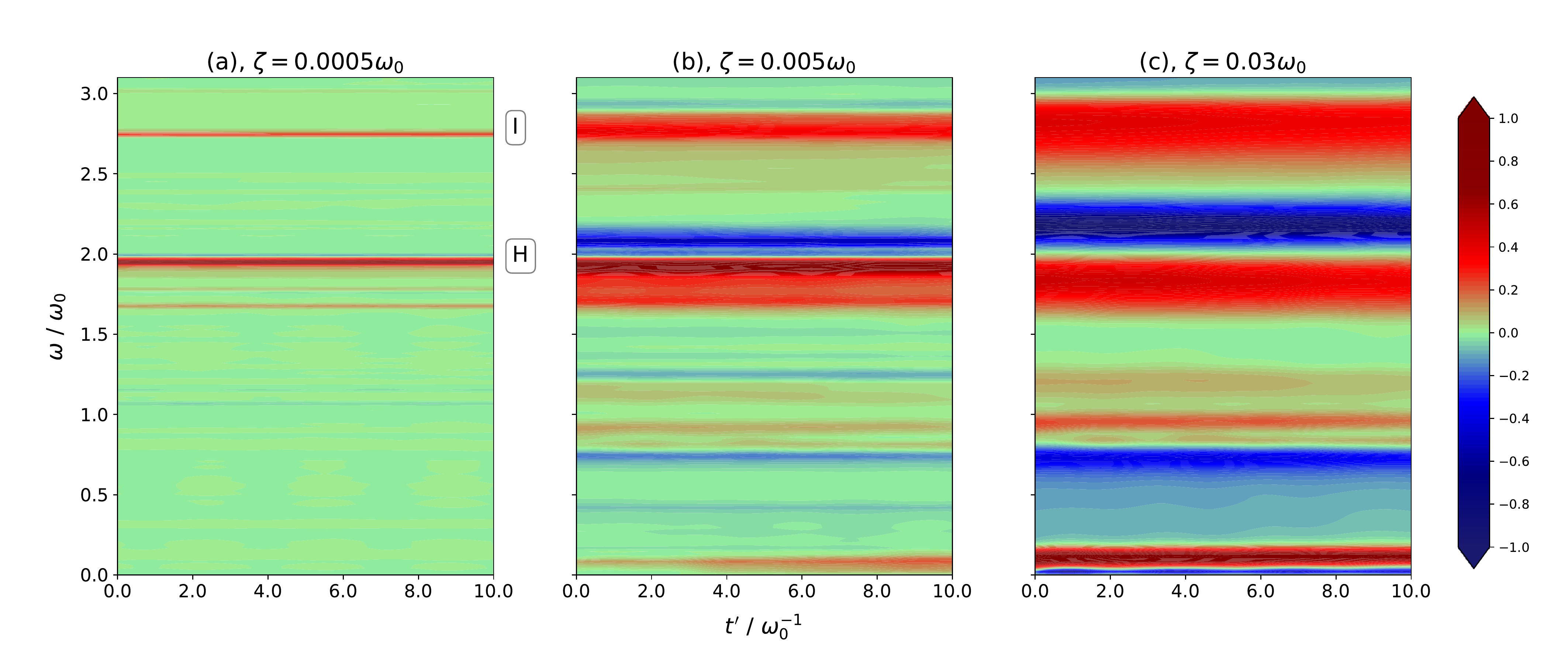}
\caption{The contour maps of TAS for the electronic excitation  ($\hat{\mu}=\hat{\mu}_e$) in the cases of the (a) weak, (b) moderate, and (c) strong system-bath interactions.
The figure is depicted in the same way with Fig. \ref{fig.transient_proton_zeta_x}.} 
\label{fig.transient_electron_zeta_x}
\end{figure}

Finally we present TAS for electronic excitation case in Fig. {\ref{fig.transient_electron_zeta_x}},  which was computed by setting $\hat{\mu}_3=\hat{\mu}_4=\hat{\mu}_e$. 
In the weak coupling case in Fig. {\hyperref[fig.transient_electron_zeta_x]{{\ref{fig.transient_electron_zeta_x}}(a)}, the peak labeled by ``H'' corresponds to the transition ${\big|\phi_{i}^{(0,n)}\big\rangle} \rightarrow {\big|\phi_{j}^{(1,n)}\big\rangle}$, and ``I'' corresponds to ${\big|\phi_{i}^{(0,n)}\big\rangle} \rightarrow {\big|\phi_{j}^{(1,n+1)}\big\rangle}$.
These two peaks represent the pure ET;  a possible transition for different $n'$ increases for larger $\zeta_x$.
In the moderate and strong coupling cases, Figs. {\hyperref[fig.transient_electron_zeta_x]{{\ref{fig.transient_electron_zeta_x}}(b)}} and {\hyperref[fig.transient_electron_zeta_x]{{\ref{fig.transient_electron_zeta_x}}}(c)}, the peaks ``H'' and ``I'' are significantly broadened and enhanced because of the constant term $V_i^{(0)}$, which introduces linear interactions between the electron subsystem and the heat bath.
Furthermore, several additional peaks appear in the range of $0.0 \le \omega_1 \le1.5 \omega_0$.
These peaks arise from the electronic transitions, but their peak locations are the same in the vibrational excitation case depicted in} Fig. {\ref{fig.transient_proton_zeta_x}}.
This result can be ascribed to the strong correlation between the electronic subsystem and the vibrational coordinates.
The increase of $\zeta$ has a promotion effect on the proton transfer, which in turn opens additional transition pathways of the electron transfer.

We also find that most of the peaks are unchanged regardless of $t^{\prime}$ even in strong coupling case. 
Thus, pure ET process is not favored in all cases for different $\zeta$ because of the pretty small electronic coupling strength $\Delta$.
With regard to the CEPT peak near $\omega = 0.05 \omega_0$, the peak intensity changes sign in both weak and strong coupling cases, as can be clearly seen from Figs. {\hyperref[fig.transient_electron_zeta_x]{{\ref{fig.transient_electron_zeta_x}}(a)}} and {\hyperref[fig.transient_electron_zeta_x]{{\ref{fig.transient_electron_zeta_x}}(c)}}.
Such variation becomes more prominent in moderate coupling case around $t^{\prime} = 2.0 \omega_0^{-1}$, if Fig. {\hyperref[fig.transient_electron_zeta_x]{{\ref{fig.transient_electron_zeta_x}}(b)}}, which indicates that a turn-over feature under a strong enough interaction occurs.
According to the above results, we find that the CEPT is the predominant process mostly because of the exact resonance conditions between initial and final states.

\begin{figure}[t]
\centering
\includegraphics[width=\textwidth]{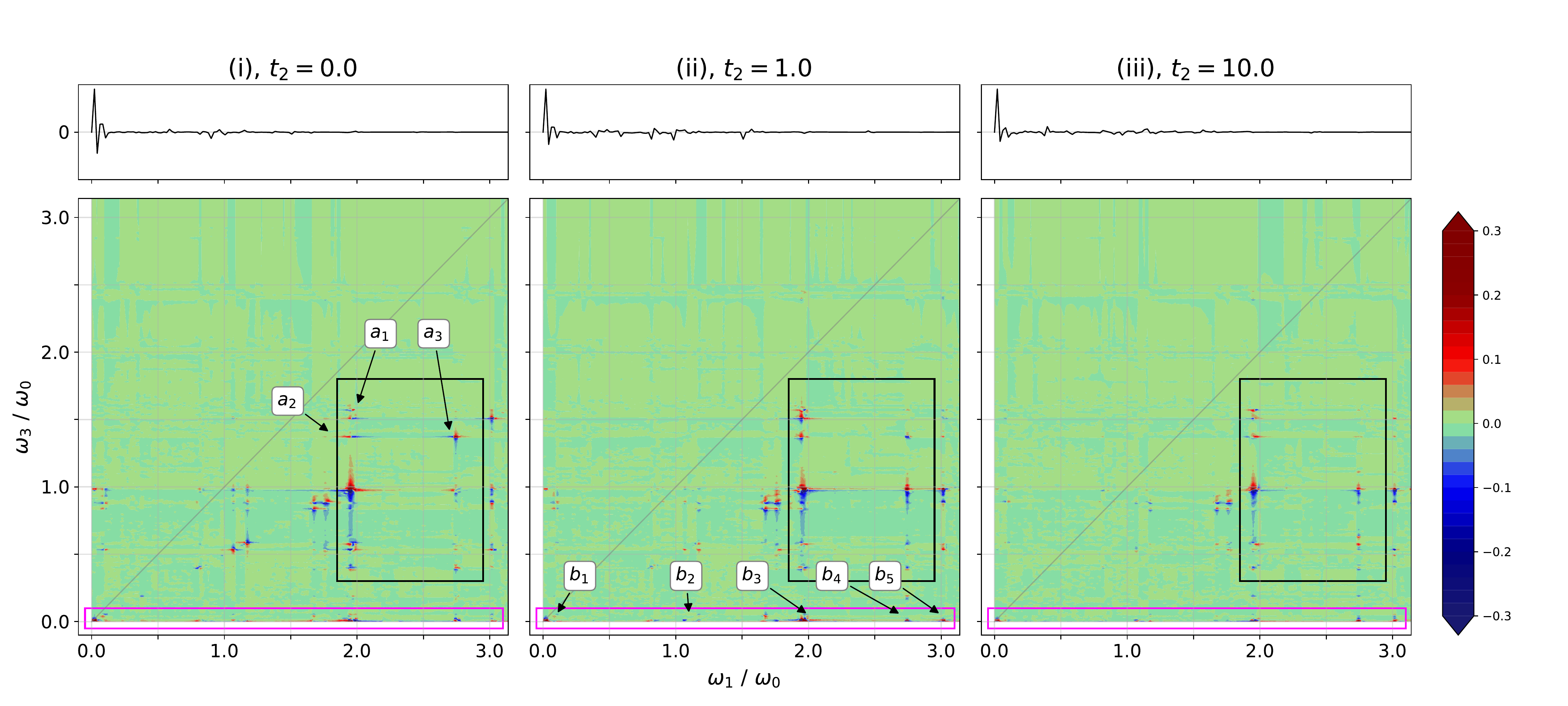}
\caption{The contour maps of 2DREVS for a weak coupling case, which correspond to the case in Figs. {\hyperref[fig.transient_proton_zeta_x]{{\ref{fig.transient_proton_zeta_x}}(a)}} and {\hyperref[fig.transient_electron_zeta_x]{{\ref{fig.transient_electron_zeta_x}}(a)}}. 
The intensities are normalized with respect to the maximum value of each case in order to see the peak profile, and the contour lines are drawn from -0.3 to 0.3. 
The red and blue curves represent the absorption and emission, respectively.
We also plot the peaks along the diagonal line in the outside above, i.e. $I(\omega_1, t_2, \omega_1)$.
The peaks in black and purple boxes represent ``ET-PT'' and ``CEPT'' peaks, while the other outside peaks are ``PT'' peaks.} 
\label{fig.2devs_weak}
\end{figure}

\subsection{Two-dimensional resonant electronic-vibrational spectroscopy (2DREVS)}

Next, we describe the 2DREVS signals as computed from Eq. {\eqref{eq.resp_5th_time}}.
The contour maps of the 2D correlation spectroscopy in the weak, moderate, and strong coupling cases are illustrated in Figs. {\ref{fig.2devs_weak}}, {\ref{fig.2devs_moderate}}, and {\ref{fig.2devs_strong}}, respectively, in which we keep the system parameters the same used to obtain the TAS signals. 
Note that most of the peaks along the diagonal line are relatively weak and not clearly visible in contour maps.
Therefore, we plot these peaks outside above as $I^{Diag}(\omega_1, t_2) = I^{Corr}(\omega_1, t_2, \omega_1)$.

The 2D correlation spectroscopy peak profiles in the weak coupling case is presented in Fig. {\ref{fig.2devs_weak}}.
For each peak, the positive intensity arises from the stimulated emission (SE) or ground state bleaching (GSB), and the negative intensity arises from the excited state absorption (ESA) for $n>0$.
Using the information obtained from TAS, we classify all the observed peaks into three parts:
1. ``ET-PT'' peaks represent the cross peaks in black boxes, which arise from the coherent ET-PT processes.
2. ``CEPT'' peaks represent the peaks in the purple box, which arise from the CEPT process. 
3. ``PT'' peaks represent the other peaks outside boxes, which are the vibrational cross peaks and represent the coherence between the proton and D-A modes.
The cross peaks for the ET process are not visible because of the small electronic coupling $\Delta$. 

We first discuss the ``ET-PT'' peaks.
Most of them appear in the same $\omega_1$ position as Fig. {\ref{fig.transient_electron_zeta_x}}, and at the same $\omega_3$ position as Fig. {\ref{fig.transient_proton_zeta_x}}, representing the corresponding ET-PT transitions. 
Here, we only concentrate on ``a1'', ``a2'' and ``a3'' that do not appear in TAS. 
In the $\omega_3$ direction, the peak ``a1'' represents the transition ${\big|\phi_{i}^{(0,n)}\big\rangle} \rightarrow {\big|\phi_{i}^{(0,n+2)}\big\rangle}$ that arises from to the SL kind of interaction in $\hat{V}_{i}$ in the $\hat{Q}$ direction.
The peaks ``a2'' and ``a3'' represent ${\big|\phi_{i}^{(1,1)}\big\rangle} \rightarrow {\big|\phi_{i}^{(0,5)}\big\rangle}$ that arise from the back PT process with a participation of the D-A mode.
These results indicate that we can analyze the combination of the ET and PT transition from the cross peaks in 2DREVS, while these contributions are mixed and appear as a single peak in TAS.

We now concentrate on the ``CEPT'' peaks.
The diagonal peak ``b1'' arises from the CEPT transition denoted as ``A'' in Fig. {\ref{fig.energy_level}}, and the other cross peaks represent the combination of CEPT-PT and CEPT-ET, where ``b2'' and ``b5'' correspond to ``A''-``D'' and ``A''-``G'', and ``b3'' and ``b4'' correspond to ``A''-``H'' and ``A''-``I'', respectively.
In addition, the peaks associated with the CEPT processes appear at symmetric positions with diagonal line.
Finally, we focus on the peak profiles at different $t_2$.
Both ``ET-PT'' and ``CEPT'' peaks stay unchanged with $t_2$ because of the weak heat-bath effect.
By contrast, the intensities of ``PT'' peaks decrease when the excited proton reaches the equilibrium distribution due to the linear interaction $V_{i,x}^{(1)}$.

\begin{figure}[h]
\centering
\includegraphics[width=\textwidth]{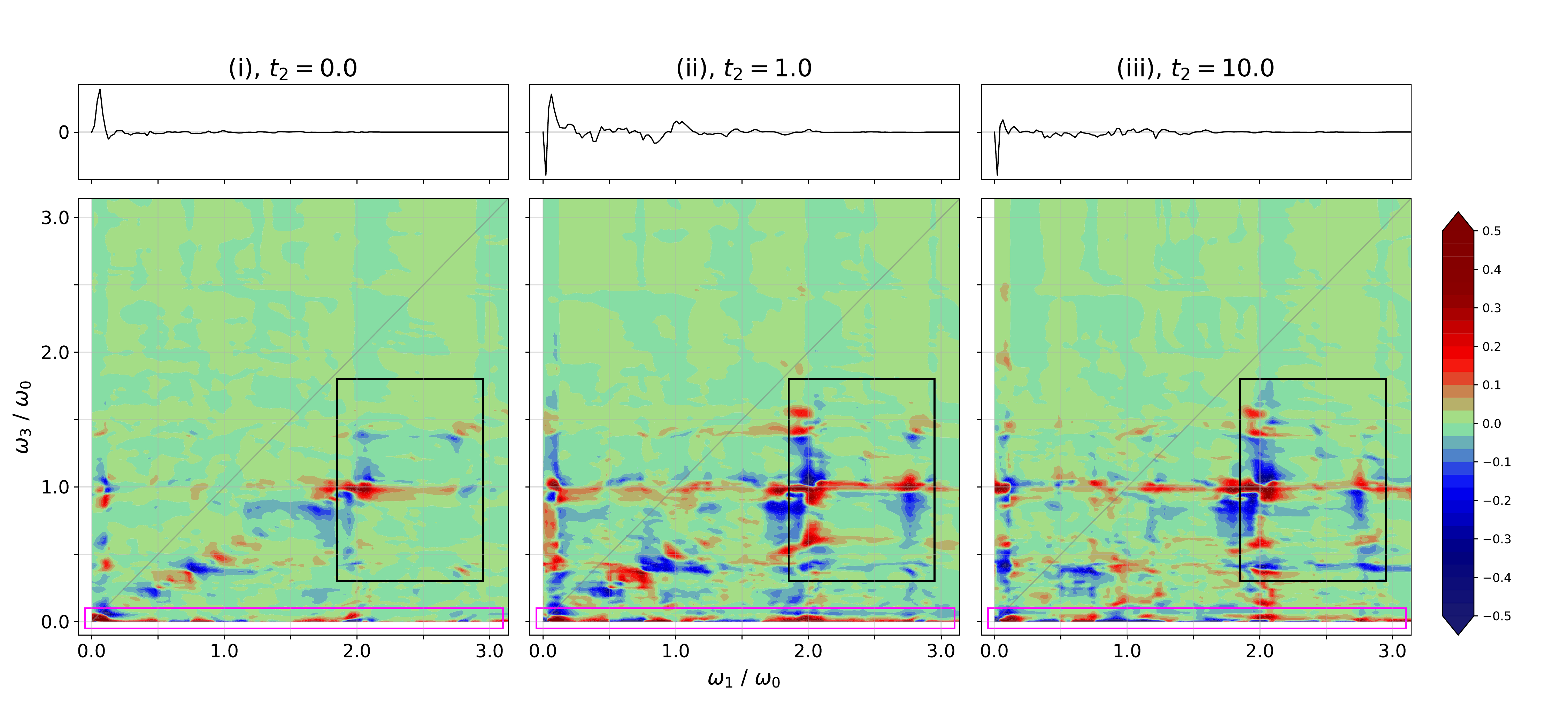}
\caption{The contour maps of 2DREVS for a moderate coupling case, which correspond to the case in Figs. {\hyperref[fig.transient_proton_zeta_x]{{\ref{fig.transient_proton_zeta_x}}(b)}} and {\hyperref[fig.transient_electron_zeta_x]{{\ref{fig.transient_electron_zeta_x}}(b)}}.
The contour lines are drawn from -0.5 to 0.5, while the other parameters are the same with Fig. {\ref{fig.2devs_weak}}.} 
\label{fig.2devs_moderate}
\end{figure}

In the moderate and strong coupling cases presented in Figs. {\ref{fig.2devs_moderate}} and {\ref{fig.2devs_strong}}, most of the peaks that are related to the proton and D-A transitions are broadened either in the $\omega_1$ or $\omega_3$ direction.
We first discuss the ``ET-PT'' peaks.
The peak positions in the $\omega_1$ direction are almost unchanged, which indicates that the system-bath interaction has a minor effect on the pure ET, as observed in TAS.
In the moderate coupling case, the peak intensities increase in Figs. {\hyperref[fig.2devs_moderate]{\ref{fig.2devs_moderate}(ii)} and {\hyperref[fig.2devs_moderate]{\ref{fig.2devs_moderate}(iii)} in comparison with Fig. {\hyperref[fig.2devs_moderate]{\ref{fig.2devs_moderate}(i)}, which indicates that the time scale of the corresponding transition is relatively short.
Most of the peaks almost disappear in the strong coupling case.
Such turn-over feature with $\zeta$ is also observed in TAS for electronic excitation, but is more clear in the 2DREVS.
This is because of the suppression effect on ET-PT coherence and mainly comes from the square linear interaction $V_{i,x}^{(2)}$.

For the ``CEPT'' peaks, the intensities increase with $t_2$, as evident from Fig. {\ref{fig.2devs_moderate}}, and become more apparent in Fig. {\ref{fig.2devs_strong}} for a larger $\zeta$.
This result indicates the existence of a bath induced CEPT process, which is also observed in TAS. 
For the ``PT'' cross peaks, most of them do not change till $t_2 = 1.0$, as illustrated in Figs. {\hyperref[fig.2devs_moderate]{\ref{fig.2devs_moderate}(ii)} and {\hyperref[fig.2devs_strong]{\ref{fig.2devs_strong}(ii)}, and almost vanish after a long $t_2$ time as illustrated in Figs. {\hyperref[fig.2devs_moderate]{\ref{fig.2devs_moderate}(iii)} and {\hyperref[fig.2devs_strong]{\ref{fig.2devs_strong}(iii)}.
Also, the intensity of the twisted positive and negative cross peak around $(\omega_1, \omega_3) = (0.8\omega_0, 0.4\omega_0)$ is reversed at $t_2 = 10.0$.
This peak mainly arises from the combination of the ``B'' and ``C'' transitions (see Fig. {\ref{fig.energy_level}}), and the reverse indicates the relaxation of excited proton. 
Thus, both proton and D-A motion have relatively longer time scales compared to CPET and ET-PT, and they are always mixed.

\begin{figure}[h]
\centering
\includegraphics[width=\textwidth]{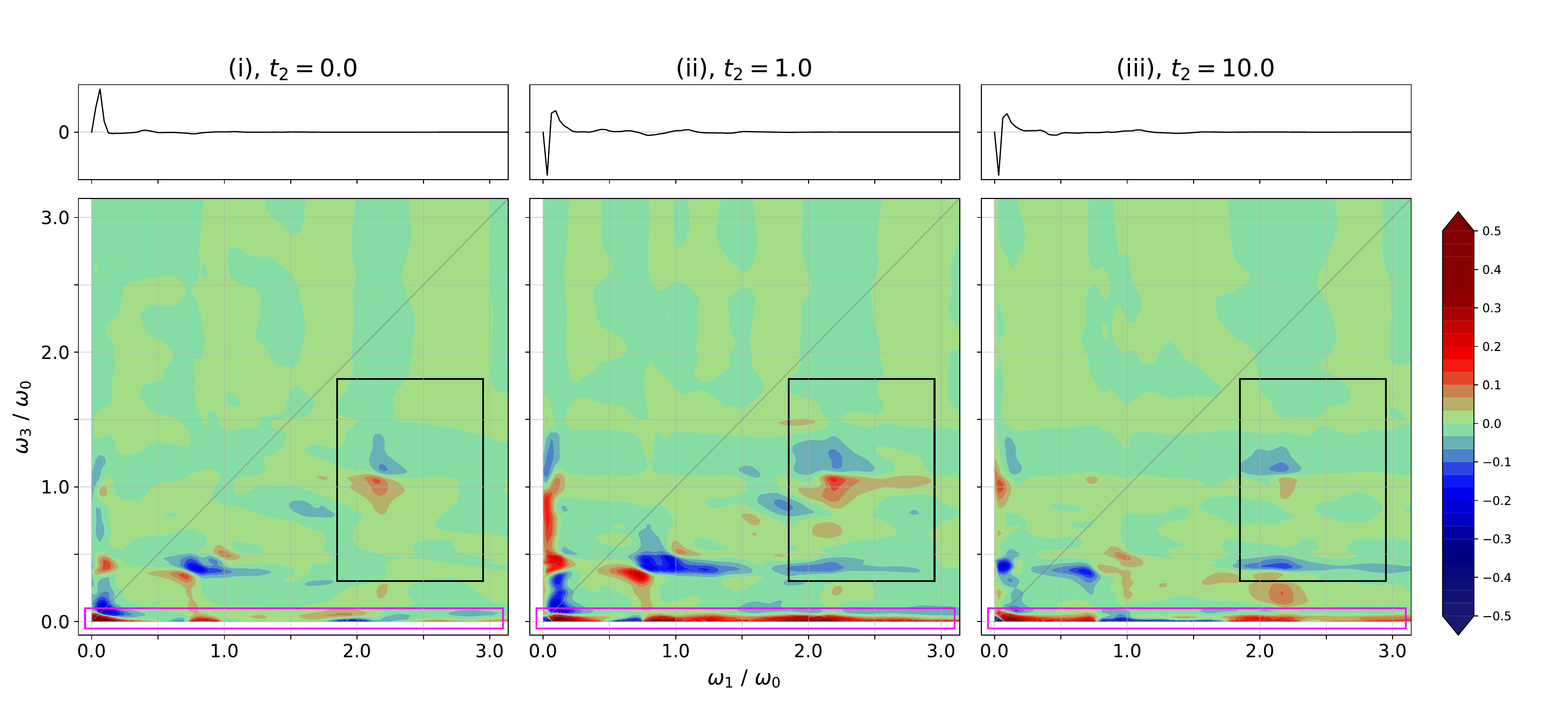}
\caption{The contour map of 2DREVS for a strong coupling case, which correspond to the case in Figs. {\hyperref[fig.transient_proton_zeta_x]{{\ref{fig.transient_proton_zeta_x}}(c)}} and {\hyperref[fig.transient_electron_zeta_x]{{\ref{fig.transient_electron_zeta_x}}(c)}}.
The contour lines are drawn from -0.5 to 0.5, while the other parameters are the same with Fig. {\ref{fig.2devs_weak}}. } 
\label{fig.2devs_strong}
\end{figure}

At the end of this section, we discuss the peaks along the diagonal line, $I^{Diag}(\omega_1, t_2)$, which represent the adiabatic transitions.
For all the coupling cases, these peaks occur in the same position found in the TAS signals, representing the corresponding transitions.
Among different $t_2$ cases, the peaks representing the CEPT transition ``A'' play a major role.
The proton and D-A mode vibrations are only visible after $t_2 = 1.0$ in Figs. {\hyperref[fig.2devs_weak]{\ref{fig.2devs_weak}(ii)} and {\hyperref[fig.2devs_weak]{\ref{fig.2devs_weak}(iii)}.
These features also corroborate the previous results that the characteristic time scale of CEPT is shorter than proton and D-A mode vibrations.
The turn-over feature is also visible because the vibration peaks become more apparent in Fig. {\hyperref[fig.2devs_moderate]{\ref{fig.2devs_moderate}(ii)} than in Fig. {\hyperref[fig.2devs_strong]{\ref{fig.2devs_strong}(ii)}.
The heat-bath plays a minor role in ET processes so that the corresponding peaks are not visible in all these cases.
Although most of the results mentioned above are also observed in TAS, 2D spectra allows a better understanding of each single contribution.

\section{Conclusion}
\label{sec.conclusion}

In this paper we introduce a system-bath model in a multi-state two-dimensional configuration space to describe the dynamics of PPCET process. 
Using the HEOM in the eigenstate representation of the system, it is possible to investigate the environment effects under a realistic system-bath interaction that causes not only fluctuation and relaxation, but also vibrational dephasing. 
Our results of population dynamics and TAS indicate that CEPT is the predominant process and has a shorter time scale when resonance conditions between initial and final states occur. 
Pure ET and PT processes also take place at much longer time.
The overall reaction would be a summation of both concerted and sequential reaction mechanism.
It is shown that  2DREVS provides a wealth of information due to the coherence among the excitation and detection periods. 
With the aids of the off-diagonal peaks, we could detect the pathway of sequential ET-PT and PT-ET transition, and concerted CEPT transition separately, whereas the diagonal peaks could reproduce the results of TAS.

Although calculating nonlinear spectra is numerically intensive, 2DREVS with TAS provides a valuable framework for studying  PPCET processes.
Since we use the eigenstate representation of the system, it is also possible to improve the description of the reacting system by increasing the dimension of its configuration space, and by introducing a more complex and structured system-bath interaction, for example, with the help of machine learning approaches.\cite{ueno2020, ueno2021}  
This provides a powerful tool to analyze the non-equilibrium reaction dynamics for rather complex PPCET reactions.

\begin{acknowledgments}
The financial support from The Kyoto University Foundation is acknowledged. 
RB acknowledges the support of the University of Torino for the local research funding Grant No. BORR-RILO-19-01.
\end{acknowledgments}

\section*{Data Availability}
The data that support the findings of this study are available from the corresponding author upon reasonable request.

\begin{appendices}
\section{Expansion of $\hat{V}_{i}$}
\label{sec.appendix.01}

In this Appendix, we expand the interaction function, $\hat{V}_{i}(\hat{x}, \hat{Q})$, with respect to $\hat{x}$ and $\hat{Q}$ up to second order as,
\begin{align}
\hat{V}_{i}(\hat{x}, \hat{Q}) =V_{i}^{(0)} + V_{i,x}^{(1)} \, \hat{x} + V_{i,Q}^{(1)} \, \hat{Q} + V_{i,x}^{(2)} \, \hat{x}^2
+ V_{i,Q}^{(2)} \, \hat{Q}^2 + V_{i,xQ}^{(2)} \, \hat{x} \hat{Q} + ...
\label{eq.V_ix_xQ}
\end{align}
where 
\begin{align}
\label{eq:V00}V_{i}^{(0)} &= \hat{V}_{i}(0, 0) 
= 2\alpha(D_i^l + D_i^r) \left(\alpha e^{\alpha x_0} - \alpha e^{2\alpha x_0}  \right), \\
\label{eq:V01}V_{i,x}^{(1)} &= \frac{\partial \hat{V}_{i}(\hat{x}, \hat{Q})}{\partial \hat{x}}\bigg|_{(0,0)} 
=2\alpha(D_i^l - D_i^r) \left(\alpha e^{\alpha x_0}- 2\alpha e^{2\alpha x_0}  \right), \\
V_{i,Q}^{(1)} &= \frac{\partial \hat{V}_{i}(\hat{x}, \hat{Q})}{\partial Q}\bigg|_{(0,0)} 
=2\alpha(D_i^l + D_i^r) \left(\alpha e^{\alpha x_0}- 2\alpha e^{2\alpha x_0}  \right), \\
V_{i,x}^{(2)} &= \frac{\partial^2 \hat{V}_{i}(x, Q)}{\partial \hat{x}^2}\bigg|_{(0,0)}  
=2\alpha(D_i^l + D_i^r)\left(\alpha^2  e^{\alpha x_0} - 4\alpha^2 e^{2\alpha x_0}  \right), \\
V_{i,Q}^{(2)} &= \frac{\partial^2 \hat{V}_{i}(\hat{x}, \hat{Q})}{\partial Q^2} \bigg|_{(0,0)} 
=\alpha(D_i^l + D_i^r)\left(\alpha^2  e^{\alpha x_0} - 4\alpha^2 e^{2\alpha x_0}  \right), \\
V_{i,xQ}^{(2)} &= \frac{\partial^2 \hat{V}_{i}(\hat{x}, \hat{Q})}{\partial \hat{x} \hat{Q}}\bigg|_{(0,0)} 
=2\alpha(D_i^l-D_i^r)\left(-\alpha^2  e^{\alpha x_0} + 4\alpha^2 e^{2\alpha x_0}  \right).
\label{eq.Vix_xQ_each}
\end{align}

\end{appendices}

\bibliography{reference}

\end{document}